\documentclass[conference, letter]{IEEEtran}
\IEEEoverridecommandlockouts
\usepackage{cite}
\usepackage{amsmath,amssymb,amsfonts}
\usepackage{algorithmic}
\usepackage{graphicx}
\usepackage{textcomp}
\usepackage{xcolor}
\usepackage[left=0.62 in,right=0.62 in,top=1.9cm]{geometry}

\usepackage{subfigure}

\def\BibTeX{{\rm B\kern-.05em{\sc i\kern-.025em b}\kern-.08em T\kern-.1667em\lower.7ex\hbox{E}\kern-.125emX}}
\begin{document}

\title{On the Performance of DCF in Full Duplex WLANs with Hidden Terminals}

\author{\IEEEauthorblockN{Anastasios C. Politis, Constantinos S. Hilas and Hristos T. Anastassiu}
\thanks{$\copyright$ 2024 IEEE. Personal use of this material is permitted. The final authenticated version is available at: https://doi.org/10.1109/BlackSeaCom54372.2022.9858233.}
\IEEEauthorblockA{\textit{Dept. of Computer, Informatics and Telecommunications Engineering} \\
\textit{International Hellenic University}\\
Serres, Greece \\
\{anpol,chilas,hristosa\}@ihu.gr}
}

\maketitle

\begin{abstract}
Full Duplex (FD) technology is considered as one of the next big leap in the evolution of modern WLANs. Allowing a node to simultaneously transmit a data frame while in receive mode, can theoretically double the system throughput. However, several requirements must be fulfilled in order for FD operation to manifest. One obvious prerequisite is that the Medium Access Control (MAC) mechanism must allow two nodes to access the shared medium simultaneously. In modern WLANs the standard MAC layer mechanism is the Distributed Coordination Function (DCF), which is specifically designed to avoid such situations. FD communications may also take place when the physical placement of the communicating parts involves the existence of hidden terminals which, in standard Half Duplex (HD) communications, imposes a significant problem. This paper investigates the performance of the Carrier Sense Multiple Access with Collision Avoidance (CSMA/CA) protocol, which constitutes the basis of the DCF mechanism, in FD WLANs with hidden terminals, and compares it with the standard HD case. Our analysis is based on performance modelling. Results indicate that, under the DCF regime, FD technology exhibits an exiguous performance improvement, in terms of saturation throughput, when compared with its half duplex counterpart. 
\end{abstract}


\section{Introduction}
The advancement of sophisticated self-interference cancellation techniques in wireless networks has opened the doors to a technology that was considered applicable only to wired networks. In-band full duplex wireless communications is now a doable venture in WLANs \cite{b1,b2,b3}. Wireless nodes in a Basic Service Set (BSS) are able to perform simultaneous transmission and reception of data frames over the same communication channel. Hence, it is not a surprise that FD communications are a strong candidate for inclusion in future IEEE 802.11 amendments \cite{b6}.

\begin{figure}[t]
  \centering
  \subfigure[Symmetric FD mode]{\includegraphics[scale=0.54]{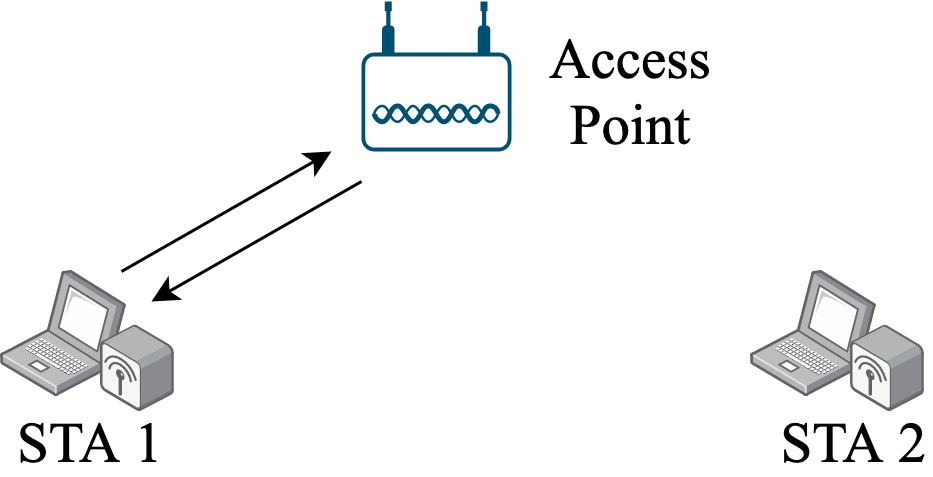}}\quad
  \subfigure[Asymmetric FD mode]{\includegraphics[scale=0.54]{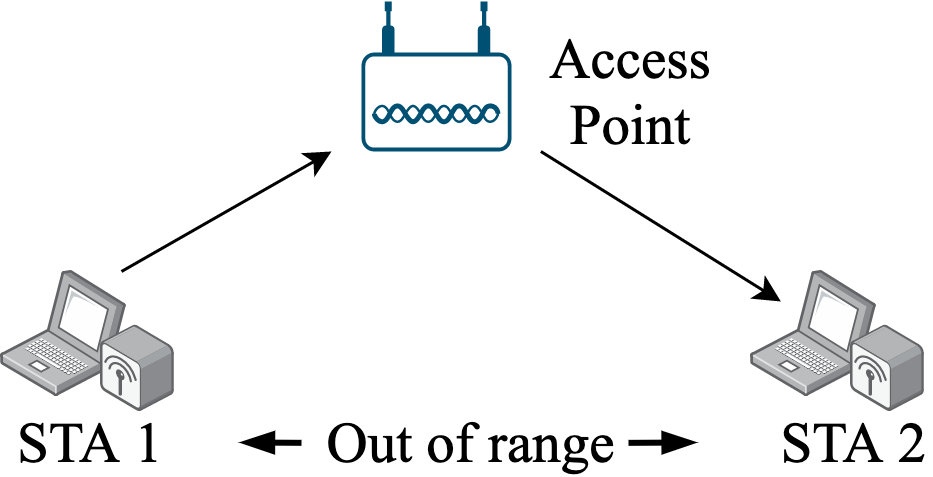}}
  \caption{FD communication modes.}
  \label{fig1}
\end{figure}

As in wired networks, FD operation in WLANs has the potential of doubling network capacity. However, unlike wired networks, this two-fold throughput improvement manifests only in two data flow configurations, which are both destination-specific. Fig.~\ref{fig1} illustrates these communication modes for a wireless BSS. Symmetric FD (SFD) mode requires the Access Point (AP) of the BSS to transmit a data frame towards a wireless station (STA) that is also transmitting back to the AP. Asymmetric FD (AFD) communication mode allows the simultaneous transmission of both an STA and the AP, but with different destinations. 

However, for FD communication to take place, a certain prerequisite, specific to the functionality of the MAC layer mechanism, must be satisfied. With regard to the current specification of the CSMA/CA protocol used today in WLANs, {\it the MAC protocol must allow the AP and exactly one STA of the BSS to access the channel simultaneously}. Technically, this means that the backoff timer of the STA and the AP must expire at the same instance. With the above condition satisfied, if the AP's transmission targets the transmitting STA, then the BSS operates in SFD mode (STAs always transmit towards the AP).
Specifically for AFD mode, a secondary condition is added: {\it the receiving STA must not be in range with the transmitting STA}, otherwise a collision event will occur. This means that the STAs must be hidden from each other in order to achieve successful AFD communications. Hence, FD operation in WLANs depends on both the underlying protocol functional details and spatial characteristics \cite{b7}. 

It becomes clear that FD operation in WLANs equipped with the DCF mechanism, as specified today, is largely opportunistic. Thus, not surprisingly, a large number of research efforts have focused on the development of appropriate MAC layer protocols suitable for FD WLANs \cite{b9,b10,b11}. However, the current DCF specification has remained virtually unchanged during the evolution of WLANs and it is unlikely that in the near future it will sustain significant modifications in a standardisation level. Hence, an evaluation of the current MAC layer protocol in FD WLANs with hidden nodes will provide a valuable performance reference point.


To this direction, the main contribution of this paper is the development of an analytical model that takes into account both FD communication modes (i.e., SFD and AFD) and considers the functionality of DCF based on its current specification (i.e., no modifications to its operation are assumed). To the best of our knowledge, similar existing models either disregard AFD communication mode or assume extensions to DCF's operation to facilitate FD transmissions. Results indicate that DCF is unable to efficiently exploit full duplex operation. More specifically, the saturation throughput improvement achieved, by employing and FD capable physical-layer (PHY) to the nodes of a BSS, turns out to be insignificant, as compared to a WLAN system operating in HD mode.



The rest of the paper is structured as follows: Section II briefly presents some of the most related research works. 
Section III analyses the spatial considerations that are needed for our performance model. Section IV describes the system model and our proposed analytical model occupies Section V. The necessary expressions to calculate the system saturation throughput are included in Section VI and the analytical results are presented in Section VII. Final remarks are provided in Section VIII. 

\section{Related Works}

In \cite{b12}, a performance analysis of CSMA/CA protocol is provided which considers the presence of hidden terminals in an FD infrastructure WLAN. However, the authors assume a constant contention window in their analysis and that the protocol provides to a receiving node (AP or STA) the ability to start a reverse direction transmission to achieve SFD communications. This assumption deviates from the actual operation of current CSMA/CA mechanism, requiring modification to its functionality. Furthermore, even though hidden nodes are included in their study, the authors focus on how the hidden node problem may be mitigated by FD technology, ignoring the possibility of AFD communications.

In \cite{b13}, a performance study of an FD BSS in non-saturated conditions is provided. Although both SFD and AFD communication modes are considered it is assumed that a node that receives a frame may initiate a secondary transmission after reading the receiving frames's header to achieve a SFD frame exchange. As before, such a capability requires a modified MAC layer mechanism.

Similarly in \cite{b14}, a performance study of CSMA/CA is provided. However, the presence of hidden terminals is not considered, thus neglecting the ability of AFD communication mode. Moreover, the study assumes that a receiving node may indirectly transmit back to the primary transmitter even if it is not in a transmission state. Again, this ability requires modifications to the MAC layer functionality.

\begin{figure}[t]
\centering{\includegraphics[width=65mm]{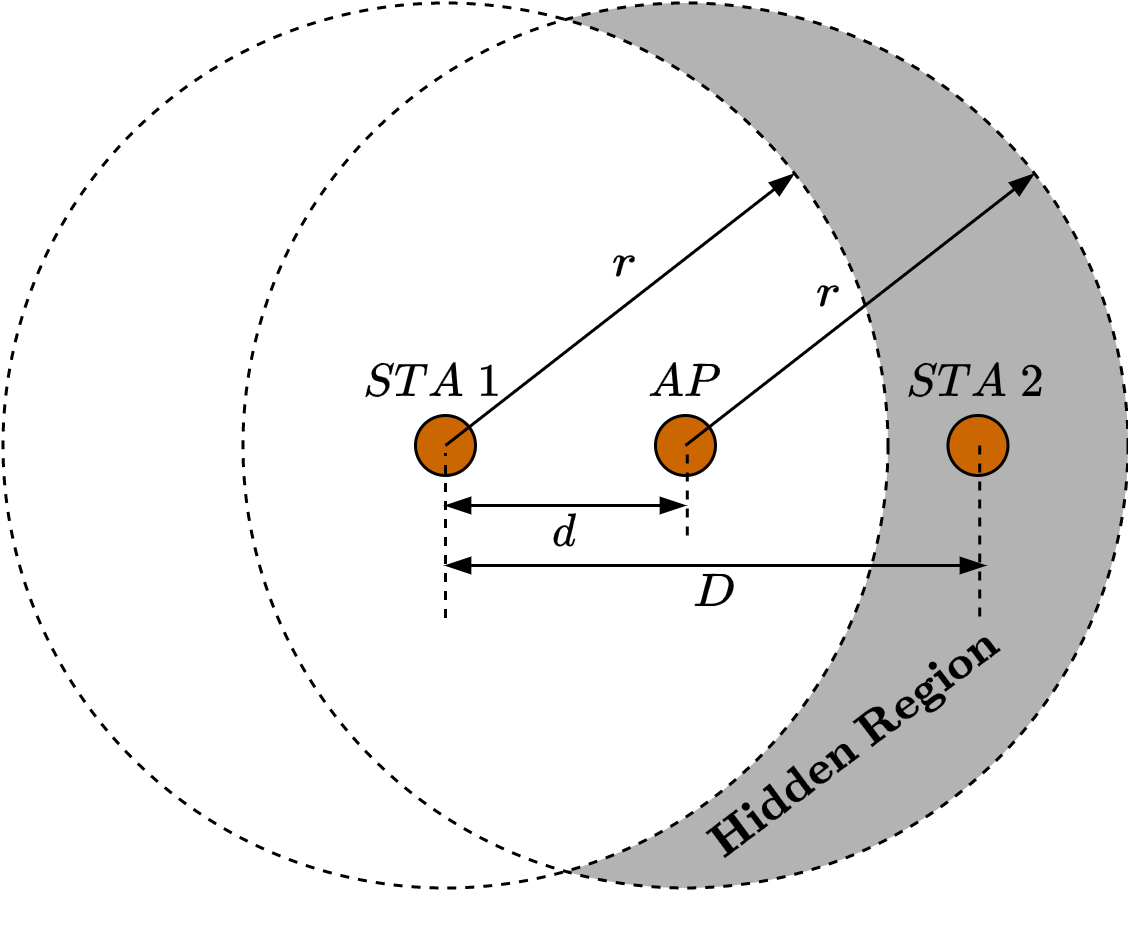}}
\caption{Illustration of the hidden region produced by the finite transmission range of wireless nodes.}
\label{fig2}
\end{figure} 

\section{Spatial Considerations}
Since SFD mode involves the AP and a single STA (which is by default within the transmission range of the AP), no spatial requirements exist. On the other hand, for AFD to be successful, the STAs engaged in this communication setting must be out of range (i.e., hidden). Hence, the physical positioning of the STAs will influence the performance of FD capable WLANs. To facilitate our analytical model, in this Section we estimate the probability that an STA is hidden from another STA in the BSS.

Fig.~\ref{fig2} re-illustrates the network of Fig.~\ref{fig1}, including geometric details regarding the transmission range of the wireless nodes. We assume that all nodes in the BSS have the same transmission range, denoted by $r$, and any transmission cannot interfere or be detected by a node positioned outside that range (i.e., detection and interference ranges are not considered). The transmission area for both STA1 and the AP are represented by two equally-sized intersecting circles whose center is separated by distance $d$ and the distance separating STA1 from STA2 is denoted by $D$. In order for STA2 to be hidden from STA1 (and vice-versa), $D$ must be greater than $r$. Hence, STA2 must be located within the shaded area depicted in Fig.~\ref{fig2}. The area of the crescent-shaped hidden region, $S_h$, produced by the equally-sized intersecting circles, is given by \cite{b15}:

\begin{equation} 
S_h=\pi r^2-2r^2\arccos\left(\frac{d}{2r}\right)+\frac{d}{2}\sqrt{4r^2-d^2}.
\label{eq1}
\end{equation}

\noindent Given that STA2 may rest anywhere within the transmission area of the AP, the geometric probability that STA2 lies within the hidden region, may be expressed as:

\begin{equation} 
p_h=\frac{S_h}{\pi r^2}.
\label{eq2}
\end{equation}

\noindent Probability $p_h$ depends on $d$ which takes values in the range $0<d \leq r$, and is maximised for $d=r$, which yields, in this case, $p_h\{max\}\approx 0.6$. For $d>r$, the STA is beyond the transmission range of the AP and as such it can not belong to the BSS.




\section{System Model}

The system model considered assumes a BSS with FD-capable wireless nodes having the same transmission range. For simplicity, carrier sensing and interference ranges are not considered. All nodes are equipped with two antennas (one for transmission and one for reception) and self-interference cancellation is presumed to be perfect. A single AP has $n$ associated STAs that are uniformly placed within the AP's transmission range. A number of hidden STAs may be present during the data transmission of another STA, hence the Request-to-send/Clear-to-Send (RTS/CTS) mechanism is assumed to precede each data frame transmission. As in \cite{b16}, the network operates in saturation conditions (i.e., the AP and all STAs always have a packet ready for transmission). Furthermore, a lossless wireless medium and a fixed frame size is assumed, and each collision leads to a complete frame loss (i.e., no capture effect). 

One key assumption in our investigation is that, {\it FD communication is possible if and only if the backoff timers of the AP and exactly one STA expire simultaneously}. With this condition, SFD and AFD communications can take place only if the AP and exactly one STA are allowed to access the channel at the same time slot. This prerequisite is needed to examine the performance of the standard DCF functionality. Unlike other performance analysis work on this topic (cf. \cite{b10}), the current MAC layer mechanism has no way to allow channel access during an ongoing transmission by another node, unless substantial modifications are made to the MAC layer protocol. 

\section{Analytical Model}

In this Section we develop an analytical model to help us asses the performance of a BSS with FD-capable wireless nodes. For this purpose, we exploit the widely endorsed work in \cite{b16}, which provides an analytical framework for estimating the performance of the standard DCF mechanism in WLANs. However, for that model to consider FD functionality, several extensions must be added. 


According to \cite{b16}, the backoff process in DCF can be modelled as a two-dimensional Markov chain. The model provides an expression for the probability that a node transmits at the beginning of a time slot:

\begin{equation} 
\tau=\frac{2(1-2p)}{(1-2p)(W+1)+pW(1-(2p)^m)}
\label{eq3}
\end{equation}

\noindent where, $W$ is the minimum contention window and $m$ is the maximum backoff stage. Probability $p$ is the conditional collision probability (i.e., the probability that, given that a node transmits, at least one other node also transmits), and is expressed as: 

\begin{equation} 
p=1-(1-\tau)^{N-1}.
\label{eq4}
\end{equation}

\noindent $N$ is the total number of nodes in the BSS (AP and STAs). Equations \eqref{eq3} and \eqref{eq4} form a nonlinear system, which can be solved using numerical methods. However, as also noted in \cite{b10}, in FD capable WLANs, the AP and STAs in a BSS are characterised by different values of $\tau$ and $p$. We denote these probabilities as  $\tau_{sta}$, $\tau_{ap}$, $p_{sta}$  and $p_{ap}$. 

Since we are concerned about the performance of the existing DCF method, the general expression in \eqref{eq3} is suitable to describe the transmission probability of a node. For the AP in the BSS we have:

\begin{equation} 
\tau_{ap}=\frac{2(1-2p_{ap})}{(1-2p_{ap})(W+1)+p_{ap}W(1-(2p_{ap})^m)}.
\label{eq5}
\end{equation}

\begin{figure}[t]
\centering{\includegraphics[width=65mm]{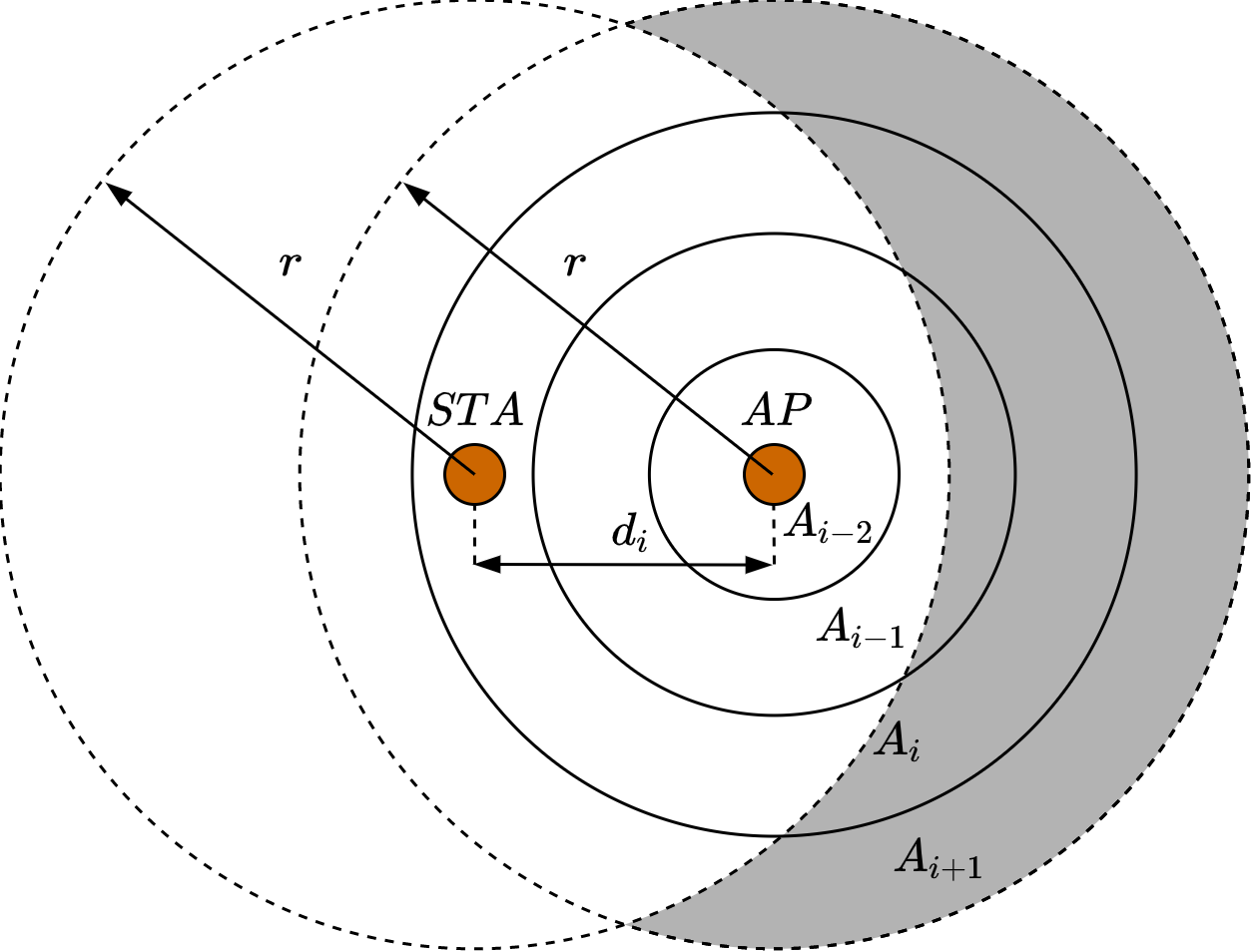}}
\caption{Division of the transmission range of a node in multiple evenly spaced concentric annuluses.}
\label{fig3}
\end{figure}

For STAs, the transmission probability, $\tau_{sta}$, is location-dependent due to the existence of hidden nodes, and as such is not the same for every STA. However, sub-groups of STAs observe the same number of hidden STAs, thus having the same $\tau_{sta}$. To support this statement, we adopt the approach followed in \cite{b18}, where the transmission range of the AP is divided into $M \in \mathbb{Z}^+$ evenly spaced concentric annuluses, $A_i$ ($i\in \{1,2,3,...,M\})$, with external and internal radii $r_i=\frac{ir}{M}$ and $r_{i-1}=\frac{(i-1)r}{M}$, respectively. This configuration is illustrated in Fig.~\ref{fig3}. 

All STAs placed on the same annulus have roughly the same distance, $d_i$, from the AP. Assuming that STAs are located at the center of the annulus's width they belong to, the distance $d_i$ of STAs in annulus $A_i$ can be estimated as:

\begin{equation} 
d_i=\frac{(i-1)r}{M}+\frac{r}{2M}=\frac{(2i-1)r}{2M}.
\label{eq6}
\end{equation}

The number of nodes that are hidden from an STA in annulus $A_i$, can be modelled as $n-1$ Bernoulli trials with $Pr[success]=p_h(i)$. Probability $p_h(i)$ is the probability that an STA lies at the hidden region produced by the STA in annulus $A_i$ and can be obtained by \eqref{eq2} for distance $d_i$. Hence, the expected number of hidden nodes perceived by each STA in $A_i$, based on the Binomial Distribution, is given by:

\begin{equation} 
h_i=(n-1)p_h(i).
\label{eq7}
\end{equation}

Supposing that the nodes are distributed proportionally to the area of each annulus, the expected number of nodes that belong to annulus $A_i$ is:

\begin{equation} 
n_i=n\frac{\pi (r_{i}^2-r_{i-1}^2)}{\pi r^2}=n\frac{i^2-(i-1)^2}{M^2}=n\frac{(2i-1)}{M^2}.
\label{eq8}
\end{equation}

\noindent In the above equation, the term $\pi(r_{i}^2-r_{i-1}^2)$ represents the area of annulus $i$. 

All $n_i$ STAs in annulus $A_i$ have the same transmission probability, which is expressed as:
\begin{align} 
\tau_{sta}(i)&=\nonumber\\
&\frac{2(1-2p_{sta}(i))}{(1-2p_{sta}(i))(W+1)+p_{sta}(i)W(1-(2p_{sta}(i))^m)}.
\label{eq9}
\end{align}

\noindent Probability $p_{sta}(i)$ is the conditional collision probability of an STA in annulus $A_i$.

Probabilities $p_{sta}(i)$ and $p_{ap}$ need to factor in the presence of hidden terminals and both SFD and AFD cases in the functionality of DCF. Thus, in the following subsections we derive suitable expressions for these probabilities.

\subsection{Conditional collision probability $p_{sta}(i)$}
 
For an STA located in annulus $A_i$, denoted as $STA_i$, a successful HD transmission is achieved when none of the remaining nodes (including the AP) attempt to transmit simultaneously (i.e., at the beginning of the same time slot). This probability is expressed as:
\begin{equation} 
\alpha(i)=(1-\tau_{ap})\pi(i)
\label{eq10}
\end{equation}

\noindent where, $\pi(i)=(1-\tau_{sta}(i))^{n_i-1}\prod_{j=1,j\neq i}^{M}(1-\tau_{sta}(j))^{n_j}$, which models the case where all remaining STAs do not transmit at the same time slot.

Furthermore, none of the $h_i$ nodes that are hidden from $STA_i$ must transmit during or before $STA_i$'s RTS frame transmission (i.e., at past or future time slots that lead to overlapping transmissions). Note that, each annulus contributes unevenly to the total number of hidden terminals, $h_i$, observed by $STA_i$. We denote the expected number of STAs in annulus $A_j$ that are hidden from $STA_i$ as $h_{i|j}$, with $j\in \{1,2,3,...,M\}$. This number is expressed as:

\begin{equation} 
h_{i|j}=n_jp_{h(i|j)}
\label{eq11}
\end{equation}

\noindent where, $p_{h(i|j)}$ is the conditional probability that an STA is hidden from $STA_i$, given that it is located in annulus $A_j$. Obviously, $\sum_{j=1}^{M}h_{i|j}=h_i$. The expression for $p_{h(i|j)}$ is derived in Appendix A. 

The duration of the RTS frame transmission can be expressed as $t_{RTS}=\rho \delta$, where $\rho$ is the number of time slots comprising the transmission duration, and $\delta$ is the propagation delay. Hence, the probability that none of the nodes that are hidden from $STA_i$ will transmit in past or future time slots that will lead to overlapped transmissions, is given by:

\begin{equation} 
\beta(i)=\prod_{j=1}^{M}(1-\tau_{sta}(j))^{h_{i|j}(2\rho-1)}.
\label{eq12}
\end{equation}

Successful transmission of $STA_i$ is also accomplished when, besides $STA_i$, only the AP initiates a simultaneous transmission towards it or towards another STA. The simultaneous transmission by the AP, regardless of its destination, will render $STA_i$'s transmission successful\footnote{If the AP transmits towards $STA_i$ then the system operates in SFD mode. However, if it targets another STA within the range of $STA_i$, the AP's transmission will fail due to interference from $STA_i$. Nevertheless, $STA_i$ will still experience a successful transmission, since a perfect self-interference cancellation is presumed. If the AP transmits towards an STA that is hidden from $STA_i$ then the system operates in AFD mode.}. Hence, this probability is:

\begin{equation} 
\gamma(i)=\tau_{ap} \pi(i).
\label{eq13}
\end{equation}

\noindent Note, that in the above case, hidden nodes do not endanger $STA_i$'s transmission, since the AP's transmission will update their Network Allocation Vector (NAV), indicating a busy medium.

Finally, $p_{sta}(i)$ can be expressed as:
\begin{align} 
p_{sta}(i)&=1-[\alpha(i)\;\beta(i)+\gamma(i)]\nonumber\\
&=\resizebox{0.4\textwidth}{!}{$1-\pi(i) \left[ (1-\tau_{ap})\prod_{j=1}^{M}(1-\tau_{sta}(j))^{h_{i|j}(2\rho-1)}+\tau_{ap} \right]$}.\nonumber\\
\label{eq14}
\end{align}

\subsection{Conditional collision probability $p_{ap}$}

Obviously, no hidden terminals exist for the AP. Hence, the AP accomplishes a successful HD transmission when none of the STAs transmit at the same time slot with the AP. This probability may be expressed as:

\begin{equation} 
\alpha(ap)=\prod_{i=1}^{M}(1-\tau_{sta}(i))^{n_i}.
\label{eq15}
\end{equation}

The AP's transmission is also successful when SFD or AFD communications take place: an STA that is the destination of AP's transmission, transmits simultaneously (SFD case), or an STA that is hidden from the AP's destination, also transmits towards the AP (AFD case). The expression for this probability is:
\begin{align} 
\beta(ap)&=\underbrace{\sum_{i=1}^{M}\frac{1}{n} \tau_{sta}(i)\pi(i)}_\text{SFD case}+\underbrace {\sum_{i=1}^{M} \frac{h_i}{n}\tau_{sta}(i)\pi(i)}_\text{AFD case}\nonumber\\
&=\sum_{i=1}^{M} \frac{h_i+1}{n}\tau_{sta}(i)\pi(i).
\label{eq16}
\end{align}

From \eqref{eq15} and \eqref{eq16} we can obtain an expression for the conditional collision probability for the AP:
\begin{align} 
p_{ap}&=1-[\underbrace{\alpha(ap)}_\text{HD case}+\underbrace{\beta(ap)}_\text{FD case}]\nonumber\\
&=1-\left[\prod_{i=1}^{M}(1-\tau_{sta}(i))^{n_i}+\sum_{i=1}^{M} \frac{h_i+1}{n}\tau_{sta}(i)\pi(i) \right].
\label{eq17}
\end{align}

\subsection{Remarks}

It is easy to prove that for $n=1$, \eqref{eq14} and \eqref{eq17} provide us with $p_{sta}(i)=0$ and $p_{ap}=0$, respectively ($\pi(i)=1$ and $h_i=0$, since $n_i=1$ and $n_j=0$). This result agrees with the intuition that for a two-node FD WLAN no collisions occur.

\begin{figure}[t]
\centering{\includegraphics[width=80mm]{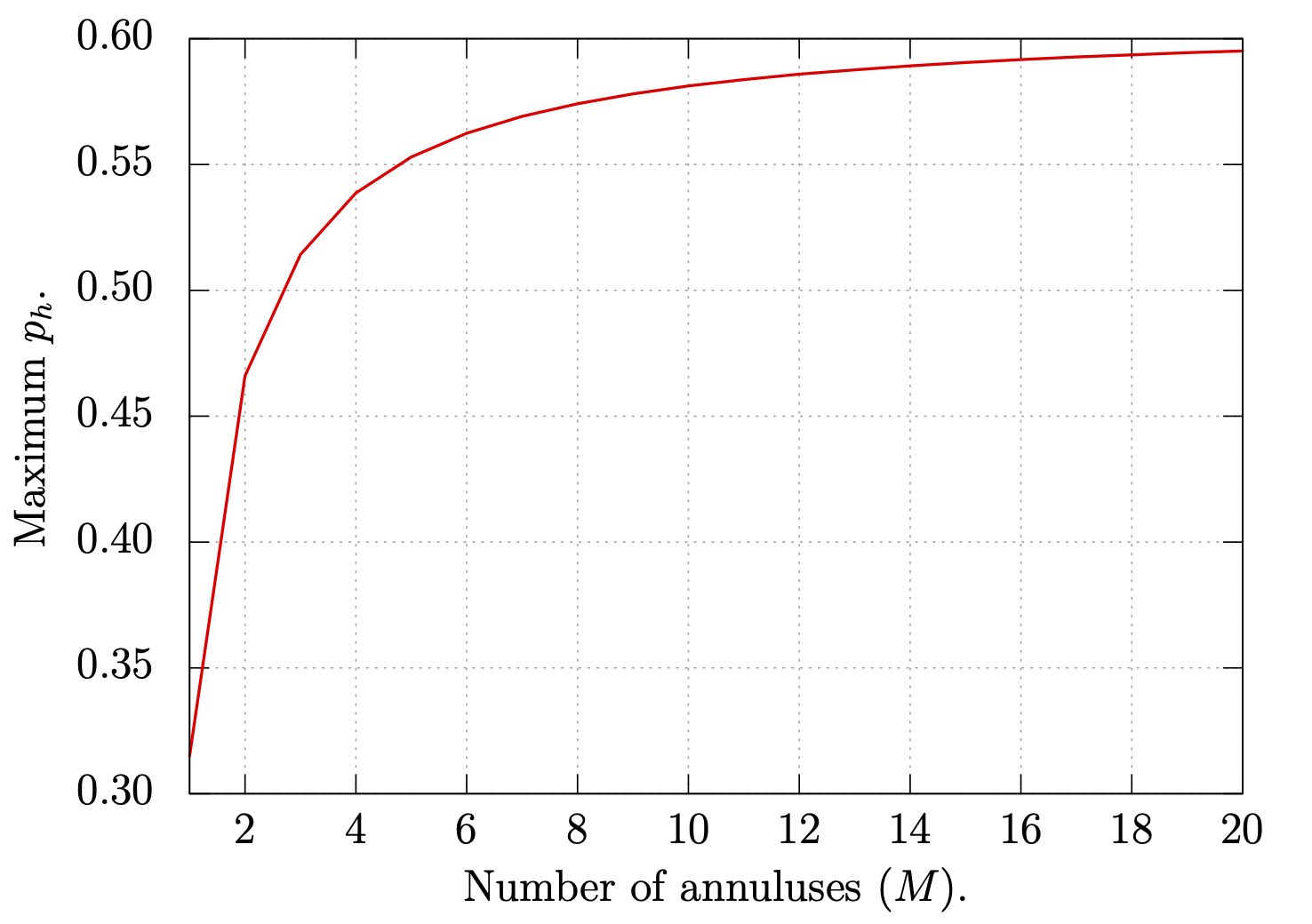}}
\caption{$p_h\{max\}$ as a function of $M$.}
\label{fig4}
\end{figure}

Equations \eqref{eq5}, \eqref{eq9}, \eqref{eq14} and \eqref{eq17}, provide $2M+2$ equations with $2M+2$ unknowns that can be solved by using numerical methods. 
Nevertheless, an appropriate value of $M$ must be selected in order to ensure an acceptable accuracy of the analytical model. Different values of $M$ will impact $p_h\{max\}$, which, as noted in Section III, reaches 0.6 for $d=r$. Fig.~\ref{fig4} shows that the computed $p_h\{max\}$ tends asymptotically to that value, as $M$ increases. The larger the value of $M$, the more accurate the analytical model will be, but at the same time the number of equations and unknowns will increase the complexity of calculations. As shown in the figure, values as low as $M=5$ can be considered acceptable. This is also supported by \cite{b18}, which concludes that fairly accurate results are produced for $M=4$. 

For $\gamma(i)=0$ and $\beta(ap)=0$, the BSS is not FD-capable and equations  \eqref{eq14} and \eqref{eq17} become:
\begin{align}
\begin{cases}
	p_{est}(i)=1-(1-\tau_{ap})\pi(i)\prod_{j=1}^{M}(1-\tau_{sta}(j))^{h_{i|j}(2\rho-1)} \\
	p_{ap}=1-\prod_{i=1}^{M}(1-\tau_{sta}(i))^{n_i}.
\end{cases}
\label{eq18}
\end{align}

\section{Saturation Throughput}

The probability that there is at least a single transmission at the beginning of a time slot can be expressed as:
\begin{equation} 
P_t=1- (1-\tau_{ap})\prod_{i=1}^{M}(1-\tau_{sta}(i))^{n_i}.
\label{eq19}
\end{equation} 

To determine the probability of successful transmission (conditioned on the fact that at least on node transmits) 
$P_s$, we must differentiate among the HD and the FD case. In the former, we end up with a successful transmission when either a single STA or only the AP transmit. Furthermore, we must consider the case where the AP and a single STA located in annulus $i$ ($STA_i$) transmit simultaneously, but the AP is targeting an STA within the transmission range of $STA_i$ (which will also lead to a successful HD transmission). These cases are included in the expression below:

\begin{align} 
P_s^{HD}&=\biggl[\tau_{ap}\prod_{i=1}^{M}(1-\tau_{sta}(i))^{n_i}+(1-\tau_{ap})\sum_{i=1}^{M}n_i\tau_{sta}(i)\pi(i)\nonumber\\
&+\tau_{ap}\sum_{i=1}^{M}\frac{n-h_i-1}{n}n_i\tau_{sta}(i)\pi(i) \biggr]\frac{1}{P_t}.
\label{eq20}
\end{align} 

A successful FD communication involves always a transmission from the AP. Thus, the probability of successful transmission for the FD case is:

\begin{equation} 
P_s^{FD}=\frac{1}{P_t}\tau_{ap}\sum_{i=1}^{M}\frac{h_i+1}{n}n_i\tau_{sta}(i)\pi(i).
\label{eq21}
\end{equation}

Finally, $P_s$ is expressed as below:

\begin{equation} 
P_s=P_s^{HD}+P_s^{FD}.
\label{eq22}
\end{equation}

The system saturation throughput ($S$) can be defined as the average amount of payload information successfully transmitted in a slot time. The mean duration of a system slot time involves an idle time slot, $\sigma$, a successful time slot, $T_{s}$, and a collision duration, $T_{c}$. Hence, $S$ is given by:

\begin{equation} 
S=\frac{P_tP_sL}{(1-P_t)\sigma +P_tP_sT_s+P_t(1-P_s)T_c}
\label{eq23}
\end{equation}

\noindent where $L$ is the packet size in bits. Parameters $T_s$ and $T_c$ can be obtained as below:

\begin{align}
\begin{cases}
	\resizebox{0.465\textwidth}{!}{$T_s=DIFS+T_{RTS}+T_{CTS}+T_{DATA}+T_{ACK}+3SIFS+4\delta$} \\
	T_c=DIFS+T_{RTS}+\delta
\end{cases}
\label{eq24}
\end{align}

\noindent where $DIFS$ and $SIFS$ are the DCF and Short Inter-Frame Spaces, respectively. $T_{RTS}$, $T_{CTS}$, $T_{DATA}$ and $T_{ACK}$ are the duration of the RTS, CTS, data and ACK frames, respectively (including PHY overhead). Propagation delay is denoted as $\delta$.

In the case of HD-capable BSS (no FD communications are feasible), \eqref{eq22} becomes:

\begin{align}
	P_s=\resizebox{0.43\textwidth}{!}{$\dfrac{1}{P_t}\bigg[\tau_{ap}\prod_{i=1}^{M}(1-\tau_{sta}(i))^{n_i}+(1-\tau_{ap})\sum_{i=1}^{M}n_i\tau_{sta}(i)\pi(i)\bigg]$}.
\label{eq25}
\end{align}

\section{Results and Analysis}

In this Section, the results of our mathematical model are presented. The model was implemented in Python programming language to obtain the numerical results. Table I summarises the PHY and MAC parameters which are used in the performance analysis.
\begin{table}[b]
\caption{PHY and MAC Parameters}
\begin{center}
\begin{tabular}{c|c}
\bf{Parameter} & \bf{Value} \\
\hline
Technology& IEEE 802.11ac  \\
\hline
MCS index& 8  \\
\hline
Spatial streams& 1  \\
\hline
Data rate& 780 $Mbps$  \\
\hline
Control rate& 6 $Mbps$  \\
\hline
PHY header duration& 44 $\mu s$  \\
\hline
MAC header length& 36 $bytes$  \\
\hline
FCS length& 4 $bytes$  \\
\hline
ACK length& 14 $bytes$  \\
\hline
RTS length& 20 $bytes$  \\
\hline
CTS length& 14 $bytes$  \\
\hline
MPDU length& 11454 $bytes$  \\
\hline
Slot duration ($\sigma$)& 9 $\mu s$  \\
\hline
Propagation delay& 1 $\mu s$  \\
\hline
DIFS& 34 $\mu s$  \\
\hline
SIFS& 16 $\mu s$  \\
\hline
Minimum contention window& 16  \\
\hline
Maximum contention window& 1024  \\
\hline
Maximum backoff stage& 6  \\

\end{tabular}
\label{tab1}
\end{center}
\end{table}

\begin{figure}[t]
\centering{\includegraphics[width=80mm]{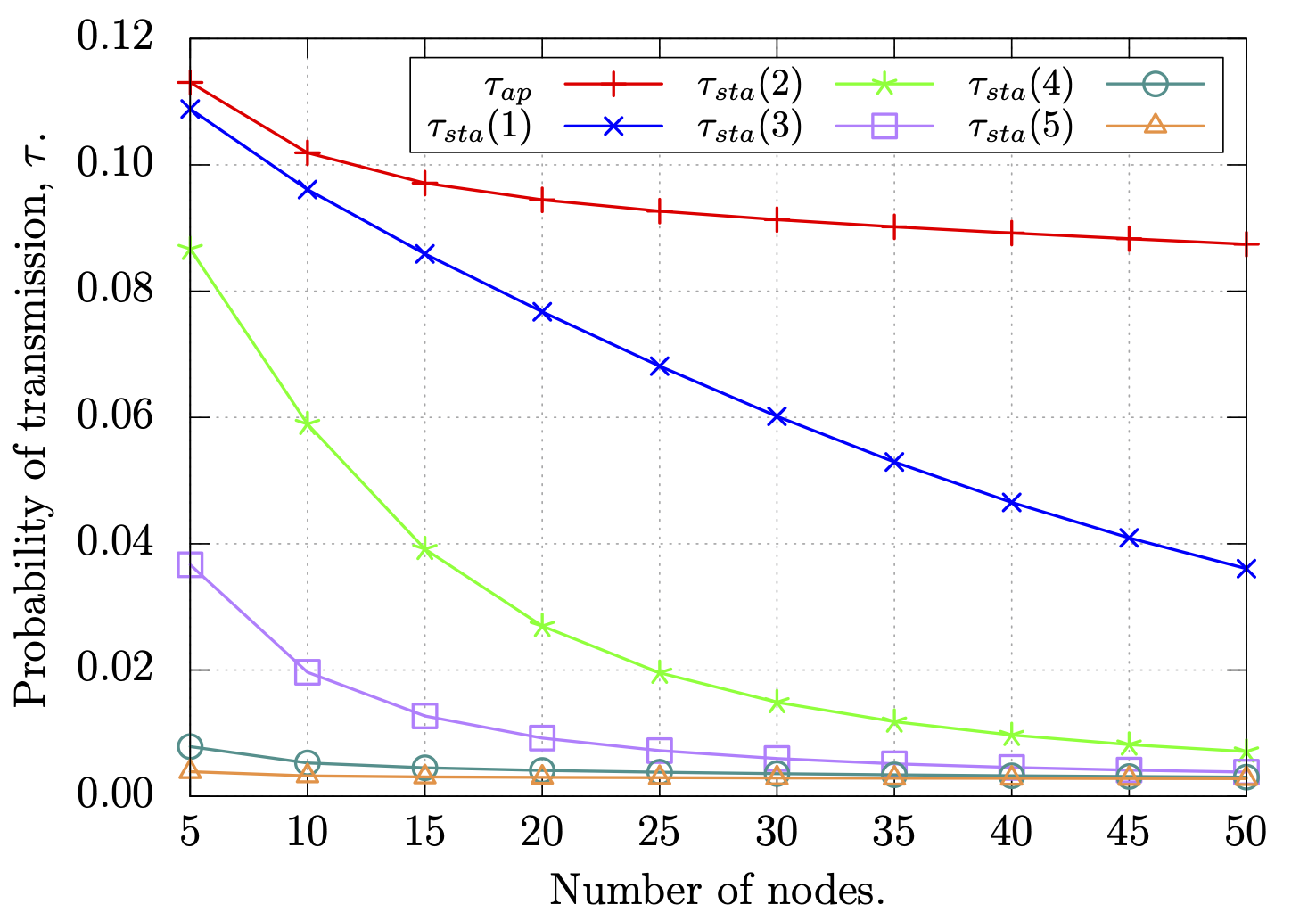}}
\caption{Numerical values of transmission probability, $\tau$, for varying number of nodes ($M=5$).}
\label{fig5}
\end{figure}

\begin{figure}[t]
\centering{\includegraphics[width=80mm]{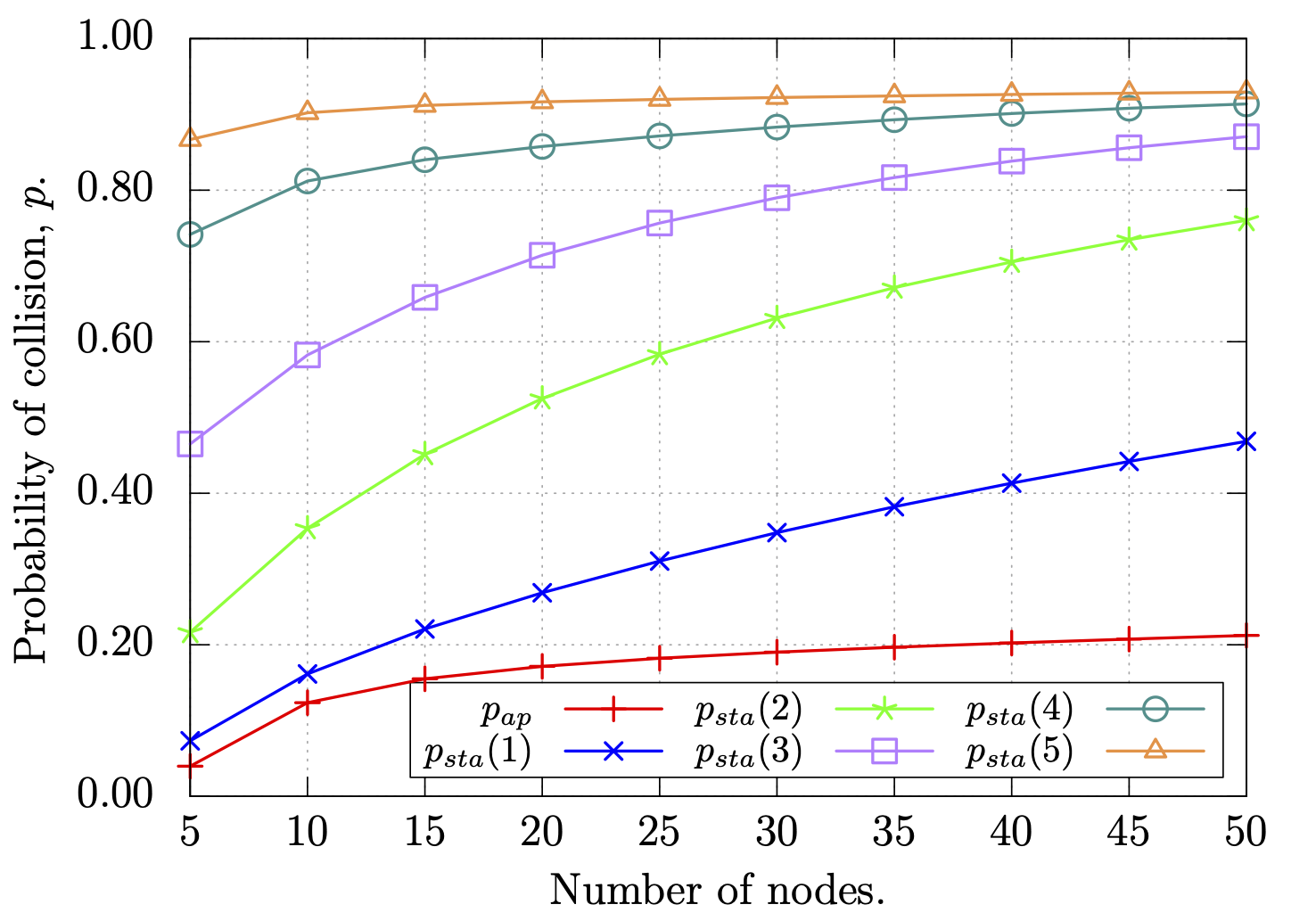}}
\caption{Numerical values of conditional collision probability, $p$, for varying number of nodes ($M=5$).}
\label{fig6}
\end{figure}

Fig. \ref{fig5} and \ref{fig6} depict the numerical values of probabilities $\tau$ and $p$, against the number of nodes participating in the BSS, respectively. The results are obtained for $M=5$. As expected, the AP exhibits the best performance in terms of $\tau$ and $p$ values, since no hidden terminals exist for that node. As far as the STAs are concerned, the closer a node is placed to the boundary of the transmission range of the AP, the less favourable its values of $\tau$ and $p$ become, since the population of hidden terminals perceived by that node scales up. 


\begin{figure}[t]
\centering{\hspace*{1cm}\includegraphics[width=80mm]{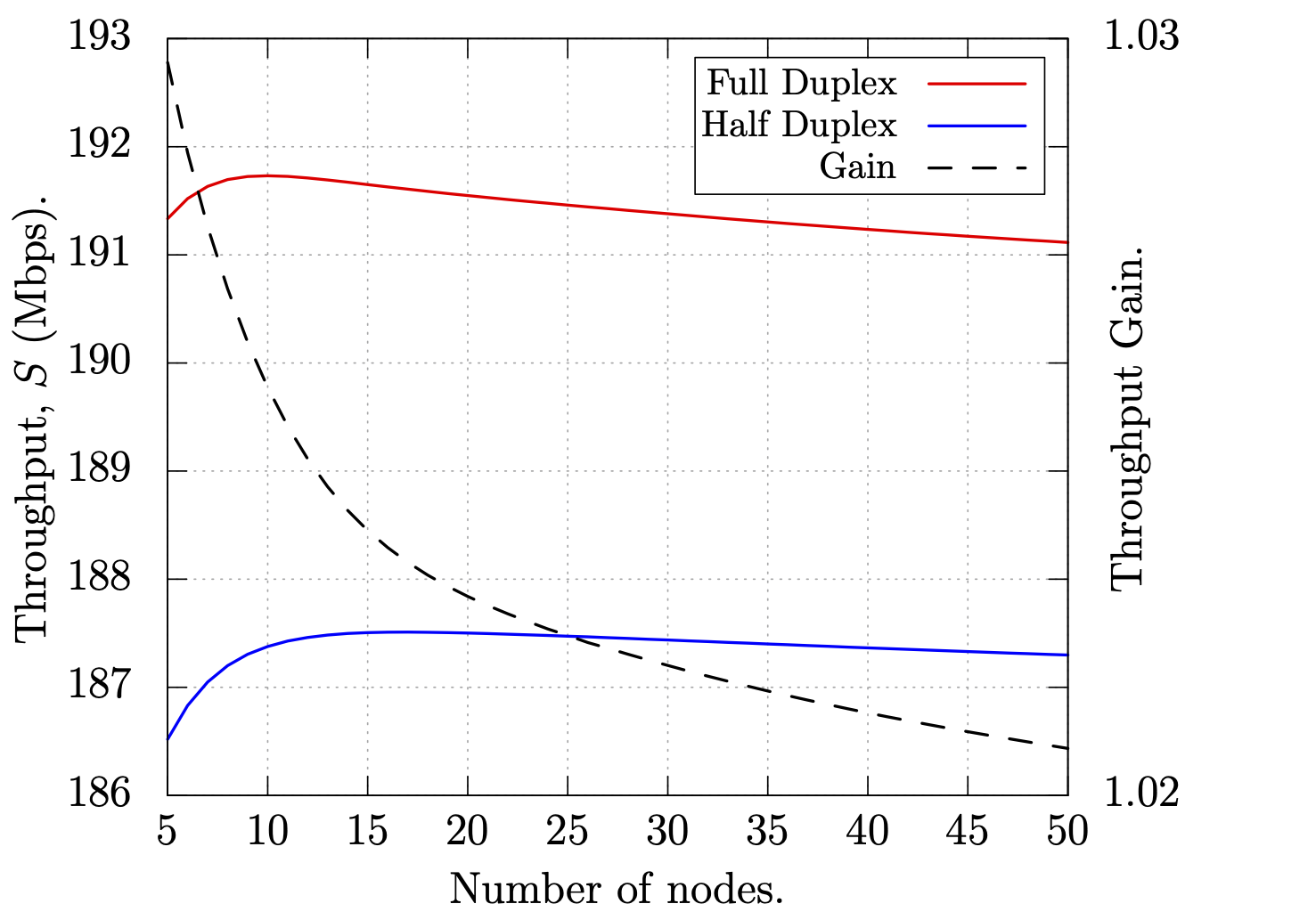}}
\caption{System saturation throughput, $S$, achieved by FD and HD cases, and throughput gain, for varying number of nodes ($M=5$).}
\label{fig7}
\end{figure}

Fig. \ref{fig7} presents the saturation throughput ($S$) for both the FD and HD cases, as well as the throughput gain obtained. It is visible that FD communications produce an inconsiderable throughput improvement, compared to the HD case. More specifically, a higher gain is observed for low number of nodes (for $n=5$ an improvement of $\approx5$ Mbps is observed, which translates to a gain of $\approx1.026$). As $n$ increases, the gain decreases, owing to the fact that the accumulating active nodes in the BSS produce more collisions and more hidden terminals (which adds to collision probability). As a result, it becomes unlikely that exactly one STA will access the medium at the same time slot with the AP, diminishing the number of SFD and AFD occurrences. It is rather more probable that, when a node wins contention (AP or STA), the remaining nodes (STAs or AP) are either at their backoff stage, or more than one of them transmit simultaneously. 

Generalising, the throughput gain that the FD capability offers over HD operation proves to be insignificant. Furthermore, the FD case tends to perform similarly with the HD case, as the number of participating nodes grows. Indeed, for a very large number of nodes ($n=1000$) we end up with $S_{FD}\approx136.252$ Mbps and $S_{HD}\approx136.219$ Mbps, which provides a throughput gain $\approx1$. The primary reason for this unfavourable behaviour roots from the fact that the current MAC layer mechanism is designed to minimise simultaneous medium access and, when this happens in a crowded BSS with hidden terminals, chances are that it will involve multiple STAs. 

\section{Conclusion}

This paper investigated the performance of the existing MAC-layer mechanism in FD WLANs with hidden terminals. To this direction, an analytical model was developed to compare FD performance with its half duplex counterpart. Results indicate that integrating FD technology in todays WLANs will not lead to any noteworthy improvement in terms of system throughput. In fact, in crowded networks with hidden nodes (which is a realistic scenario), no gain is to be expected by applying the FD technology. The main reason for this unsatisfactory performance is that the current MAC layer mechanism is specifically engineered to avoid simultaneous transmissions, which is the basic prerequisite for FD communication mode to manifest. This strengthens the necessity of a complete re-design of the MAC-layer access method in order to showcase the benefits of full duplex communications in modern WLANs. 

\appendices
\section{Derivation of Conditional Probability $p_{h(i|j)}$}

\renewcommand{\theequation}{\thesection.\arabic{equation}}
\setcounter{equation}{0}

Referring to Fig.~\ref{fig3}, we focus on a particular annulus $A_j$ with $n_j$ STAs ($j\in \{1,2,3,...,M\}$). To calculate the probability that a single STA in $A_j$ lies also in the hidden region produced by the STA located at $A_i$, $p_{i|j}$, an expression for the area of the intersection between two circular disks is needed. If the disks have radii $r$ (radius of STA's transmission range) and $r_j$ (external radius of $A_j$), and the distance of their centers is $d_i$, then this area is given by \cite{b15}:
\begin{align} 
&X(r,r_j,d_i)=r_j^2\arccos\left(\frac{d_i^2+r_j^2-r^2}{2d_ir_j}\right)\nonumber\\
&+r^2\arccos\left(\frac{d_i^2+r^2-r_j^2}{2d_ir}\right)\nonumber\\
&-\frac{\sqrt{(-d_i+r_j+r)(d_i+r_j-r)(d_i-r_j+r)(d_i+r_j+r)}}{2}.
\label{eq26}
\end{align}

\noindent Note that for $r_j=r$ the above equation reduces to \eqref{eq1}.
The area of the $r_j$-disk not covered by the $r$-disk (i.e., the part of the $r_j$-disk that belongs to the hidden region) is equal to:
\begin{align} 
Y(r,r_j,d_i)=\pi r_j^2-X(r,r_j,d_i).
\label{eq27}
\end{align}

\noindent For $r_j$-disks that are covered by the $r$-disk (i.e., no part of the $r_j$-disk is in the hidden region produced by the $r$-disk), $Y(r,r_j,d_i)=0$. 
The part of annulus $A_j$ that belongs to the hidden region equals to:
\begin{align} 
Z(r,r_j,r_{j-1},d_i)=Y(r,r_j,d_i)-Y(r,r_{j-1},d_i)\geq 0.
\label{eq28}
\end{align}

Now, probability $p_{i|j}$, can be expressed as:
\begin{align} 
p_{h(i|j)}=\frac{Z(r,r_j,r_{j-1},d_i)}{\pi (r_{j}^2-r_{j-1}^2)}.
\label{eq29}
\end{align}


\section*{Acknowledgment}

The authors wish to acknowledge financial support by the Special Account for Research Funds of the International Hellenic University (former Technological Educational Institute of Central Macedonia), Greece, under grant SAT/IE/93/6/28-2-2019/50135.

\end{document}